\documentclass[a4paper,12pt]{article}
\pdfoutput=1 

\usepackage{jcappub} 
\usepackage[utf8]{inputenc}
\usepackage{graphics}
\usepackage{epstopdf}
\usepackage{hyperref}
\usepackage{amsmath}
\usepackage{relsize}
\usepackage{psfrag}
\usepackage{epsfig}
\usepackage{xcolor}
\usepackage{url}

\usepackage[T1]{fontenc} 
\usepackage[percent]{overpic}
 \normalsize
\newcommand{\bmat}{\left(\begin{array}}
\newcommand{\emat}{\end{array}\right)}
\newcommand{\be}{\begin{equation}}
\newcommand{\ee}{\end{equation}}
\newcommand{\bea}{\begin{eqnarray}}
\newcommand{\eea}{\end{eqnarray}}

\usepackage{booktabs}
\usepackage{siunitx}
\def\lsim{\raise0.3ex\hbox{$\;<$\kern-0.75em\raise-1.1ex\hbox{$\sim\;$}}}
\def\gsim{\raise0.3ex\hbox{$\;>$\kern-0.75em\raise-1.1ex\hbox{$\sim\;$}}}

\begin{document}
\begin{flushright}
IPPP/22/52
\end{flushright}
\vspace{-0.9cm}
\title{Cosmological Imprints of SUSY Breaking in Models of Sgoldstinoless Non-Oscillatory Inflation}
\author{L. Heurtier$^{1}$, A. Moursy$^{2}$ , and L. Wacquez$^{1,3}$ }
\affiliation{
$^1$Institute for Particle Physics Phenomenology, Durham University, South Road, Durham, U.K.\\
$^2$Department of Basic Sciences, Faculty of Computers and Artificial Intelligence, Cairo University, Giza 12613, Egypt.\\
$^3$Universit\'e Paris-Saclay, 91405, Orsay, France}
\emailAdd{lucien.heurtier@durham.ac.uk}

\emailAdd{a.moursy@fci-cu.edu.eg}

\emailAdd{lucien.wacquez@universite-paris-saclay.fr}
\vspace{2pt}
\abstract{
In supergravity, the dynamics of the sgoldstino -- superpartner of the goldstino superfield associated with the breaking of supersymmetry at low energy -- can substantially modify the dynamics of inflation in the primordial Universe. So-called sgoldstinoless models assume the existence of a nilpotency constraint $S^2=0$ that effectively removes the sgoldstino from the theory. Such models were  proposed to realise non-oscillatory inflation scenarios with a single scalar field, which feature a long period of kination at the end of inflation, and therefore a non-standard post-inflationary cosmology. Using effective operators, we propose models in which the sgoldstino is stabilized close to the origin to reproduce the nilpotent constraint. We show that small sgoldstino fluctuations may lead to a sizeable back-reaction on the cosmological history. We study the effect of this back-reaction on the inflation observables measured in the cosmic microwave background and confront the model to a series of constraints including limits on $\Delta N_{\rm eff}$. We show that the peculiar form of the potential in the large supersymmetry breaking scale limit can generate peaks in the scalar power spectrum produced from inflation. We study how certain perturbation modes may re-enter the horizon during or after kination and show that a large supersymmetry breaking scale may lead to the formation of primordial black holes with various masses in the early Universe.}
\maketitle
\flushbottom

\section{Introduction}
\label{sec:intro}

In 1973, Volkov and Akulov understood that at energies below the scale of supersymmetry breaking $f_0$, supersymmetry (SUSY) is non-linearly realized \cite{Volkov:1973ix}. Whereas such realization of SUSY can be understood from a purely geometrical point of view \cite{Volkov:1973ix, Ivanov:1978mx}, it is also possible to recover non-linear SUSY from a linear realization of SUSY, by imposing certain multiplicative constraints on the superfields of the theory, such as a nilpotency constraint on the Goldstino superfield, $S^2=0$ \cite{Rocek:1978nb, Casalbuoni:1988xh, Komargodski:2009rz}. Effectively, the effect of such constraints is to impose the decoupling of heavy states below the scale of supersymmetry breaking, reproducing nonlinear supersymmetry in an effective low-energy actions. In the context of scalar-field inflation, such constraints represent a powerful tool to write down simple scalar theories which can be used to describe early cosmology in simple terms \cite{Antoniadis:2014oya, Ferrara:2014kva, Kallosh:2014via, Kallosh:2014hxa, Linde:2015uga}. In the context of string theory, the emergence of such constraints has been extensively studied \cite{Aparicio:2015psl,McGuirk:2012sb,Kallosh:2014wsa,Bergshoeff:2015jxa,Kallosh:2015nia,Bandos:2015xnf,Garcia-Etxebarria:2015lif,Dudas:2000nv,Antoniadis:1999xk,Sugimoto:1999tx}. Nevertheless, it is generically difficult in such UV embeddings to restore linear SUSY above the scale $f_0$, meaning that the scalar component of the goldstino $S$ satisfying the constraint $S^2=0$ simply does not exist. From a dimension-four quantum field theory point of view, the essence of such constraints remains obscure. The authors of Ref.~\cite{Komargodski:2009rz, DallAgata:2016syy} showed that sgoldstino-less theories, and more generally theories with constrained superfields, can be derived as an effective field theory, by using large non-renormalizable operators. However, it was shown in Refs.~\cite{Buchmuller:2014pla,Buchmuller:2015oma,Argurio:2017joe} that the large Wilson coefficients limit required in order to obtain an exact nilpotency constraint is in tension with the idea that such operators could originate from an effective field theory. As a matter of fact, the excursions of the sgoldstino to non-zero field values during inflation can have a drastic impact on the dynamics of inflation and on its cosmological observables \cite{Dudas:2016eej, Argurio:2017joe}. 

In the vast majority of inflation models, the inflaton oscillates around the minimum of its potential after inflation, and eventually decays into Standard Model (SM) particles, leading to the hot big bang phase of the Universe's history. However, there exists a subclass of models in which the inflaton ends up rolling along a very flat trajectory after inflation ends. This eternal rolling of the inflaton field, which eventually enters a regime of slow-roll due to Hubble friction, may provide an explanation to the nature of dark energy in so-called {\em quintessential inflation} models \cite{Akrami:2017cir, Dimopoulos:2022rdp,deHaro:2021swo,AresteSalo:2021wgb,Akrami:2020zxw,Agarwal:2018obp,Dimopoulos:2017zvq,WaliHossain:2014usl,Bastero-Gil:2009wdy,Neupane:2007mu,Rosenfeld:2005mt,Campos:2004xw,Dimopoulos:2001ix,Dimopoulos:2001qu,Yahiro:2001uh,Dimopoulos:2000md,Peloso:1999dm,Peebles:1998qn}. More generally, models of {\em non-oscillatory} inflation consider the possibility that the inflaton may simply not oscillate around a minimum at the end of inflation, but, instead, keeps rolling along a flat direction \cite{Ellis:2020krl,Campos:2004xw,Feng:2002nb}.

Sgoldstino-less models of $\alpha$-attractors   reveal to be extremely useful to construct inflationary models in SUGRA which can roll along a flat direction after the end of inflation, and where supersymmetry is naturally broken during and after inflation\cite{Carrasco:2015rva, Carrasco:2015pla, Akrami:2017cir}. However, obtaining the nilpotency constraint on the goldstino superfield from an effective field theory description in the context of no-oscillatory inflation models remains, up to now unexplored. In this paper, we aim to construct effective field theories of supergravity, in which the sgoldstino is naturally stabilized, and where non-oscillatory inflation can be successfully realized. Interestingly, a general feature of non-oscillatory inflation is that it features a long period of {\em kination}, an era during which the energy density is dominated by the kinetic energy of the inflaton. As we will see, the existence of such an era, together with the peculiar dynamics of the inflaton field in the presence of SUSY breaking will lead to interesting cosmological signatures, which could act as smoking gun evidences of such models in cosmological data.

The paper is organized as follows: In Sec.~\ref{sec:SUSYB-QI} we introduce the framework in which we will be working throughout the paper. We present in particular two models which are thoroughly studied throughout the paper. In Sec.~\ref{sec:inflationobservables} we exhibit the predictions of those models regarding the perturbation modes in the CMB and confront the model to several constraints, including limits on $\Delta N_{\rm eff}$. Finally, in Secs.~\ref{subsec:pbh} we show that the model can predict the existence of primordial black holes which may account for the full relic density of dark matter or simply perturb the post-inflationary history of the Universe by dominating the energy density before evaporating.

\section{Supergravity Model of Non-Oscillatory Inflation with SUSY Breaking}
\label{sec:SUSYB-QI}
In order to easily build inflation potentials in the presence of a nilpotency constraint $S^2=0$ for the goldstino superfield, we use the framework developed in Ref.~\cite{DallAgata:2014qsj}. 

%
We parametrize our lagrangian as
\bea\label{eq:lagrangian}
W &=& \left(1- \dfrac{S}{\sqrt{3}}  \right)^3 f(Z)\\
K & = &  K_1 \left(  Z , \overline{Z}  \right)   - 3 \log\left[  1- \dfrac{|S|^2}{3} +  \dfrac{|S|^4}{\Lambda^2}  \right]\,,
\eea
where $f(Z)$ is a holomorphic function that satisfies \cite{DallAgata:2014qsj}
\be \label{eq:condf}
\overline{f(Z)}= f(- \, \overline{Z})\,,  
\hspace{0.5cm} f(0)\neq 0\,,
\ee
and where the chiral superfield $S$ denotes the goldstino superfield, responsible for the breaking of SUSY \cite{DallAgata:2014qsj}. The chiral superfield $Z$ contains the inflaton and the UV scale $\Lambda$ plays the role of stabilizing the sgoldstino during the whole history of the Universe, ensuring that the constraint $S^2=0$ is approximately satisfied. Throughout the paper we work in units of reduced Planck mass, where $M_p=1$.

In the models we consider in the next sections, we checked both analytically and numerically that the fields Re$(Z)$ and Im$(S)$ have masses larger than the Hubble scale and remain stabilized to zero at all time of the field evolution. Therefore, we can safely restrict our study to the dynamics of the two remaining real degrees of freedom, Im$(Z)=z$ and Re$(S)=s$.  Following Ref.~\cite{DallAgata:2014qsj}, Eq. (\ref{eq:condf}) implies that
\be
\overline{f'}\left( - i \, z \right)  = - f'\left( i \, z \right)\,\,,\hspace{0.5cm}\overline{f}\left(- i \, z \right)  =  f\left( i \, z \right)\,.
\ee

In the purely nilpotent limit, the stabilizing term $|S|^4/\Lambda^2$ is sent to infinity, while the sgoldstino $s$ is assumed to vanish at all time. We will denote by $V_{\rm nil}(z)$ the potential obtained in this limit. During inflation, SUSY is broken in both the sgoldstino direction and the inflaton direction, whereas after inflation ends, SUSY will be broken mainly along the $S$ direction. The gravitino mass squared is then given by
\be 
m_{3/2}^2\simeq \left|f\left( i \, z \right) \right|^2= f\left( i \, z \right) ^2
\ee 
After one relaxes the pure $S^2=0$ constraint, it was shown in Ref.~\cite{Dudas:2016eej} that $s$ can acquire a non-zero vacuum expectation value during inflation which may affect the inflationary dynamics. The value of the sgoldstino vev $\langle s\rangle$ can be expressed as a function the purely nilpotent potential $V_{\text{nil}}(z)$ as \cite{Dudas:2016eej}
\bea\label{eq:svev}
\langle s \rangle 
= \frac{\Lambda ^2 \, V_{\text{nil}}}{2 \sqrt{3} \left[\Lambda ^2 \,  V_{\text{nil}}+ 6 \, m_{3/2}^2\right]}\,.
\eea
The scalar potential then acquires a correction from the non-zero vev of $s$ during the dynamics of inflation and late time expansion, as follows
\bea
V(z,s)= V_{\text{nil}}(z) + s V_1 (z)+ s^2 V_2(z) +\hdots
\eea 
where we have expanded up to the second order in small $s$, and $V_j(z)$ are given by
\bea
V_1(z) &=& - 2\sqrt{3} V_{\text{nil}}(z) \,, \\
V_2(z) &=&  \frac{36  \, m_{3/2}^2+ 6 \, \Lambda ^2 \,  V_{\text{nil}}}{ \, \Lambda ^2}\,,
\eea
and the sgoldstino mass squared is $m_s^2 = V_2$. Using the non-zero vev of $s$ (\ref{eq:svev}), the inflaton effective potential takes the form
\bea 
V_{\text{eff}}(z)= V_{\text{nil}}(z) - \dfrac{\Lambda^2 \, V_{\text{nil}}(z)^2 }{2 \left[ 6 m_{3/2}^2 +  \Lambda^2 V_{\text{nil}}(z) \right] }\,.
\eea
%

\subsection{Model I- alpha-attractors}
\label{sec:modelI}
\begin{figure}[h!]
	\centering
	\includegraphics[width=0.48\textwidth]{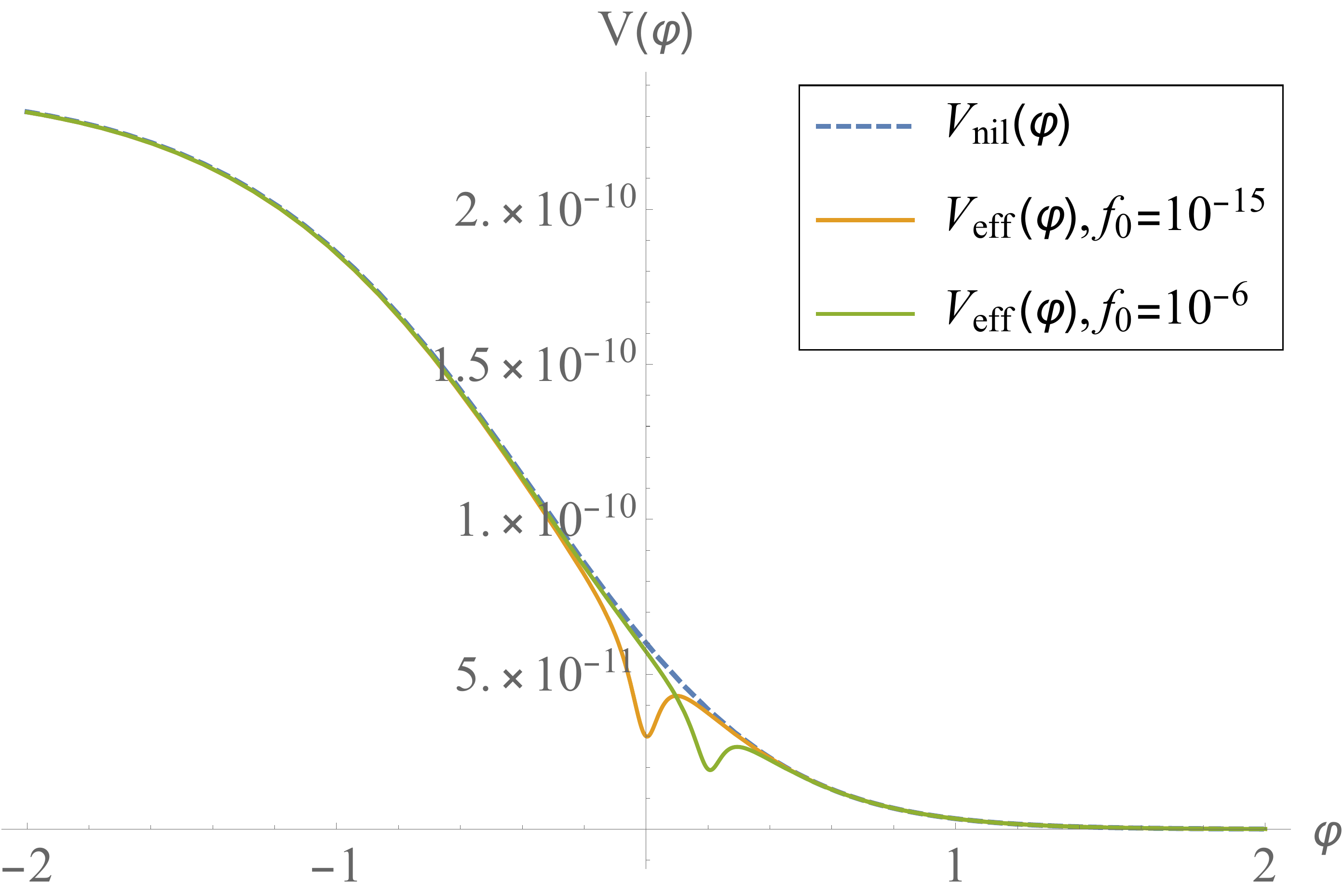}
	\hspace{0.1cm}
	\includegraphics[width=0.48\textwidth]{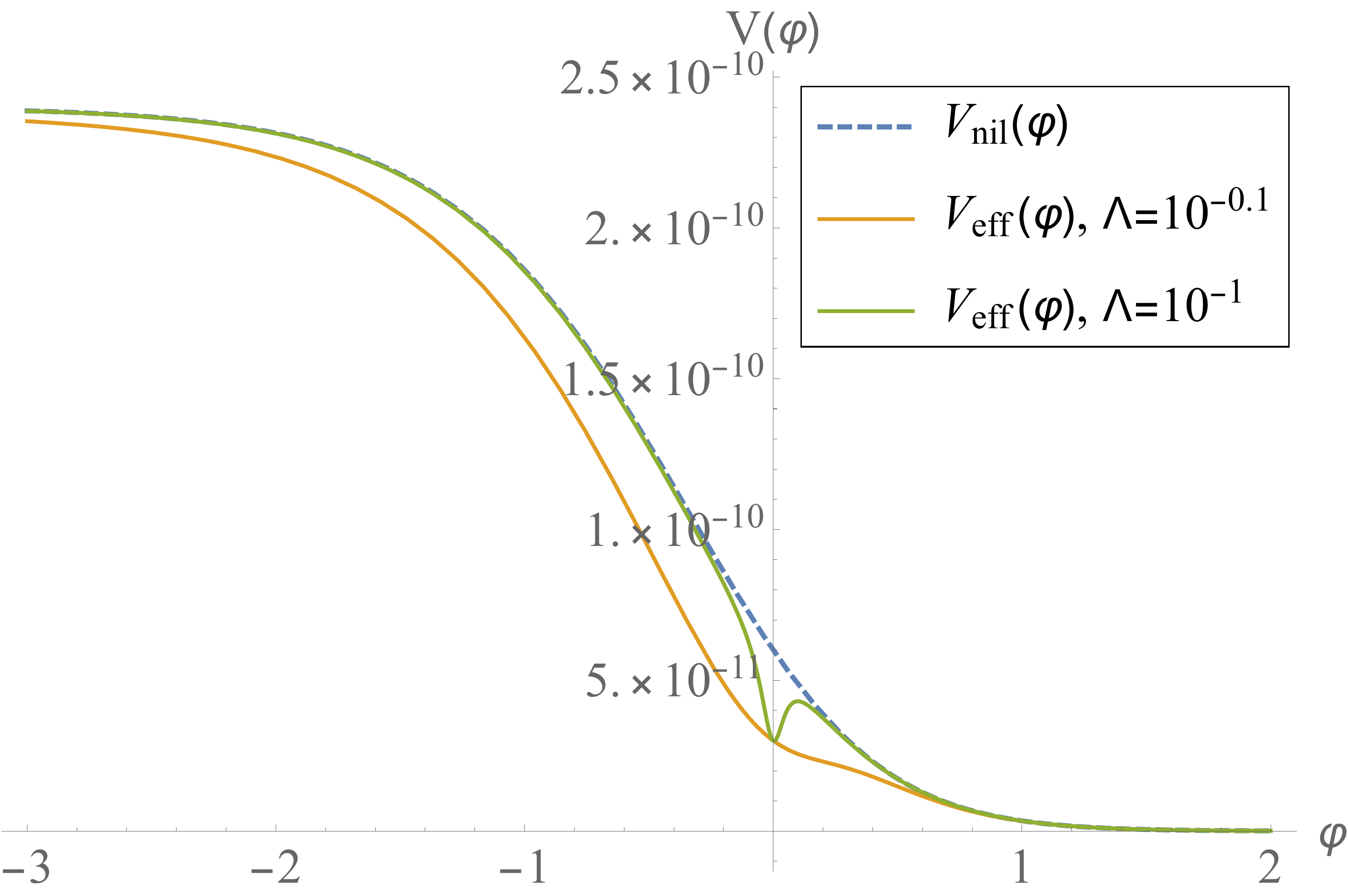}
	\caption{The effective scalar potential of model I, $V_{\text{eff}}(\varphi)$, in orange and green colors with $V_0=1.5\times10^{-12}, \, \alpha=1/6, \, $ and changing $f_0$ with fixing $\Lambda= 0.1$ in the left panel, while  changing $\Lambda$ and fixing $f_0=10^{-15}$ in the right panel. The blue dashed curve corresponds to the nilpotent limit. All values are given in the units where $M_P=1$.
	\label{fig:potI}}
\end{figure}

The first model we consider is an $\alpha$-attractor, which can be realized by taking the K\"ahler potential to be of the form 
\bea\label{eq:modelI}
K_1 =   - \dfrac{3\alpha}{2} \log\left[  \dfrac{(1-|Z |^2)}{(1+Z^2)\,(1+ \overline{Z}^2)}   \right] \,.
\eea
In this case, the purely nilpotent potential can be computed using the $s=0$ limit:
\be 
V_{\text{nil}}= \dfrac{  \left(1-z^2 \right)^2}{3\alpha} \left|f'\left( i \, z \right) \right|^2=
-\dfrac{  \left(1-z^2 \right)^2}{3\alpha} \left[f'\left( i \, z \right)\right] ^2\,.
\ee
In order to obtain a potential which can describe a non-oscillatory inflation dynamics, we consider the function $f(Z)$ to be 
\be 
f(Z)=f_0 - \sqrt{ V_0}  \log(1-i \, Z)\,.
\ee
Note that this choice is not unique. However, as we will see in what follows, the qualitative consequences which arise from the dynamical behaviour of the sgoldstino along the cosmological timeline are relatively generic. Thanks to the set up proposed in Ref.~\cite{DallAgata:2014qsj} that we have chosen here, it is remarkable that the scale of inflation $V_0$ is disconnected from the scale of SUSY breaking at low energy $f_0$ since the inflation potential will be driven mainly by $f'(iz)$, as we will see shortly. In order to normalize the inflaton field, we perform the following redefinition
\be
z = \tanh\left[ \dfrac{\varphi}{\sqrt{6\alpha}} \right] \,.
\ee
Therefore, in the nilpotent limit, the scalar potential takes the form
\be 
V_{\text{nil}}= \frac{ V_0}{3}  \left[\tanh \left(\frac{\varphi }{\sqrt{6 \alpha }}\right)-1\right]^2 = \frac{4 V_0}{3 \alpha  \left(e^{\sqrt{\frac{2}{3 \alpha }} \varphi }+1\right)^2},
\ee
which features two different plateaus with a hierarchical difference of the order $V_0$ as can be seen from Fig. \ref{fig:potI}. Note that in this case, the plateau which lies at $\varphi\to \infty$ is at zero, meaning that another mechanism should be invoked in order to uplift the potential to $\Lambda\sim 10^{-120}M_p^4$ and address the cosmological constant problem. We do not address this issue in this paper and instead assume that some other mechanism will be responsible for the value of the cosmological constant today. We instead focus on the post-inflationary cosmology which takes place since the inflationary era until matter-radiation equality. Nevertheless, it can be noted that a {\em cosmological seesaw} mechanism is operating in our model which naturally sources a large energy density $H^2\sim V_0/3$ during inflation and suppresses the potential energy of the inflationary sector when $\varphi\to 0$. After relaxing the exact nilpotency condition $S^2=0$ and keep a finite scale $\Lambda$ in the lagrangian, the scalar potential acquires a correction due to the non-zero vev of $s$ and the effective potential has the form
%
\bea
V_{\text{eff}}(\varphi)&=& \frac{4 V_0}{3 \alpha  \left(e^{\sqrt{\frac{2}{3 \alpha }} \varphi }+1\right)^2} \nonumber \\
 && - \frac{4 \Lambda ^2 V_0^2}{27 \alpha^2  \left(e^{\sqrt{\frac{2}{3 \alpha }} \varphi }+1\right)^4 \left[\left(f_0-\sqrt{V_0} \log \left(\dfrac{2}{e^{-\sqrt{\frac{2}{3 \alpha }} \varphi }+1} \right)\right)^2+\frac{2 \Lambda ^2 V_0}{9 \alpha  \left(e^{\sqrt{\frac{2}{3 \alpha }} \varphi }+1\right)^2}\right]}\,. \nonumber\\
\eea

\subsection{Model II- canonical K\"ahler}
\label{sec:modelII}
\begin{figure}[h!]
	\centering
	\includegraphics[width=0.48\textwidth]{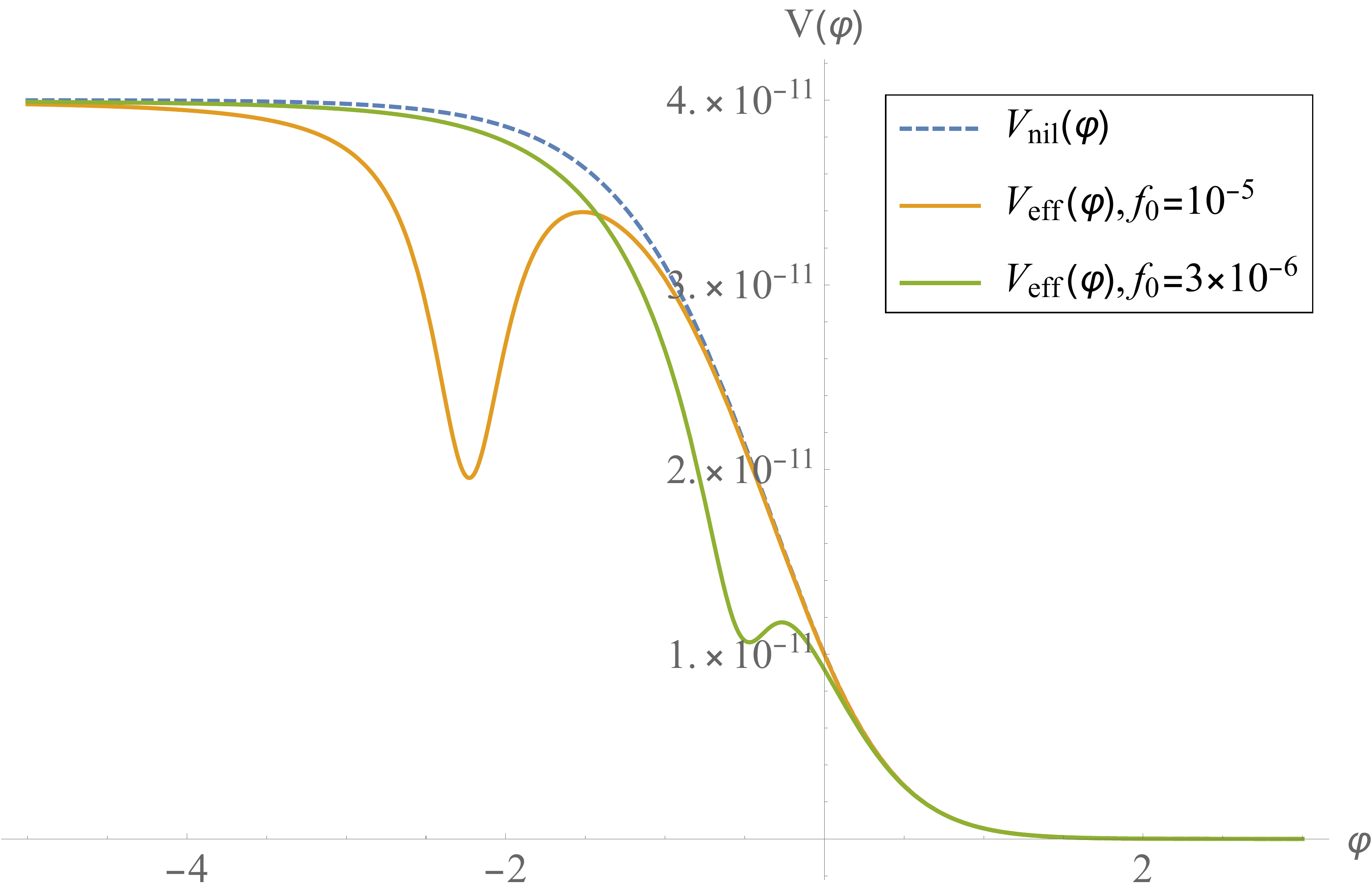}
	\hspace{0.1cm}
	\includegraphics[width=0.48\textwidth]{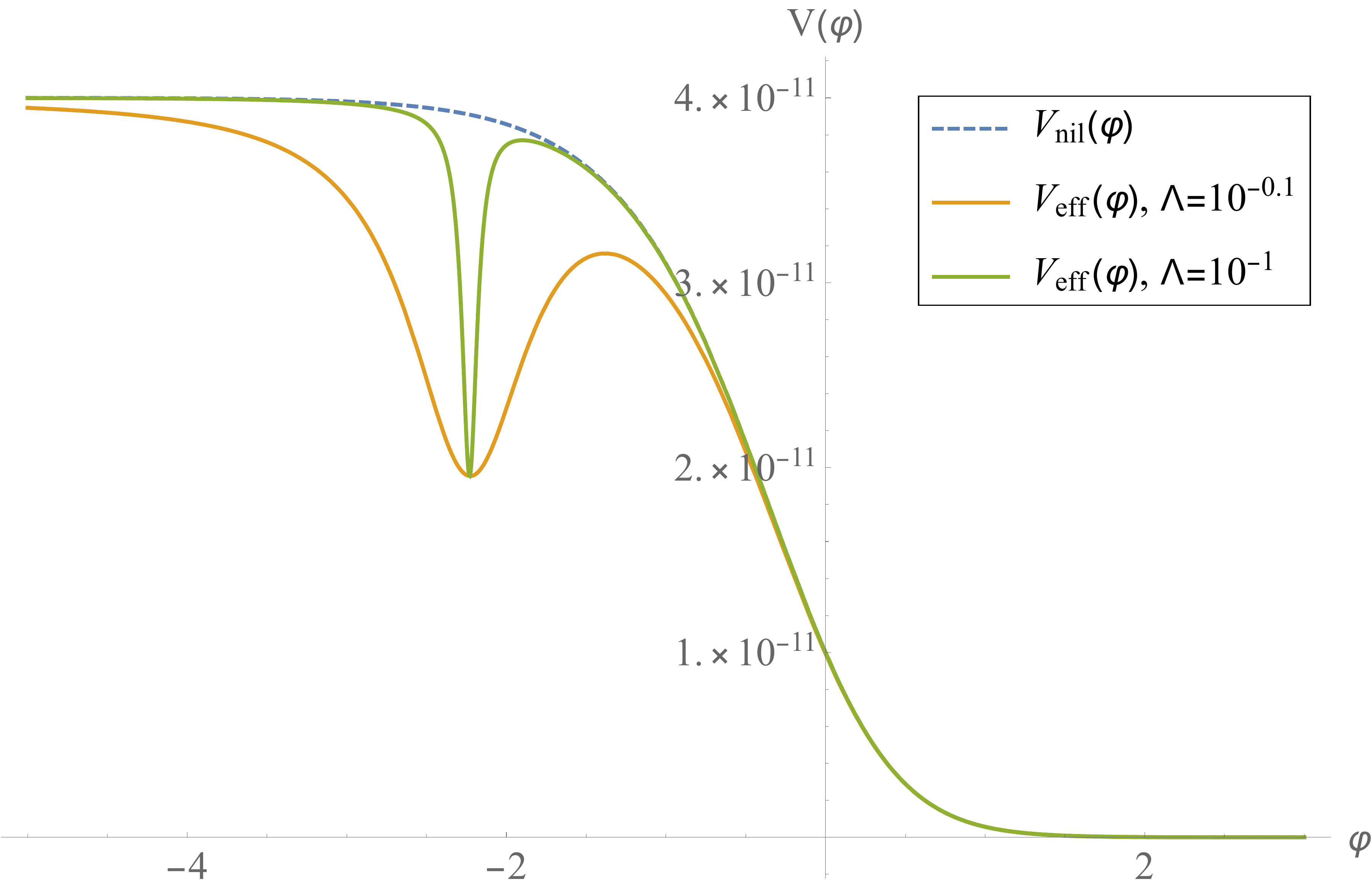}
	\caption{The effective scalar potential of model II,  $V_{\text{eff}}(\varphi)$, in orange and green colors with $V_0=5\times10^{-12},  \,b=1 $ and changing $f_0$ with fixing $\Lambda= 0.5$ in the left panel, while  changing $\Lambda$ and fixing $f_0=10^{-5}$ in the right panel. The blue dashed curve corresponds to the nilpotent limit. All values are given in the units where $M_P=1$.
	\label{fig:potII}}
\end{figure}

The second model that we consider simply features a canonical K\"ahler potential where we used a shift symmetry is applied in order to to avoid the $\eta$-problem
\bea\label{eq:modelII}
K_1 & = &  \dfrac{1}{2} \left(  Z + \overline{Z}  \right)^2  .
\eea
Repeating the procedure described above, in the exact nilpotent limit the inflation potential has the form
\be 
V_{\text{nil}}=  \left|f'\left( i z \right) \right|^2=
- f'\left( iz \right) ^2,
\ee
We then choose the function $f(Z)$ to be
\be 
f(Z)=f_0 - \sqrt{ V_0}  \log\left(1+e^{2\sqrt{2} \, i \, b \, Z} \right) \,.
\ee
Again $V_0$ represents the inflation scale while the SUSY breaking scale is given by $f_0$, b is a dimensionless parameter, and we normalize the field redefining $z=\dfrac{\varphi}{\sqrt{2}}$.
In the nilpotent limit, the scalar potential becomes
\be 
V_{\text{nil}}= \frac{8\, b^2 V_0}{\left(e^{2 b \varphi }+1\right)^2}\,,
\ee
and has again two plateaus, one at the inflation scale $V_0$ and the other at 0, as depicted in Fig.~\ref{fig:potII}. After the sgoldstino dynamics is turned on, the effective potential becomes
%
\bea
\!\!\!\!\!\!\!\!\!\!\!\!\!\!\! V_{\text{eff}}(\varphi)&=& \frac{8 \, b^2 V_0}{\left(e^{2 b \varphi }+1\right)^2}  
 - \frac{16 \, b^4 \Lambda ^2 V_0^2}{\left(e^{2 b \varphi }+1\right)^4 \left[\dfrac{4 b^2 V_0 \Lambda ^2 }{\left(e^{2 b \varphi }+1\right)^2}+3 \left(f_0-\sqrt{V_0} \log \left(e^{-2 b \varphi }+1\right)\right)^2\right]} \nonumber\\
\eea 
%

\section{From Cosmic Inflation to the Present Universe}
\label{sec:inflationobservables}
In models of oscillatory inflation, the inflaton is expected to decay into SM particles after it oscillates at the bottom of its potential. In that case, the dynamics of the inflaton scalar field after the time of cosmic reheating may not leave any imprint in the later history of the Universe and can be safely ignored once the Universe has entered its hot big-bang phase. In the case of non-oscillatory inflation, it is extremely important to track the dynamics of the scalar field throughout the whole history of the Universe, as it can contribute significantly to the total energy density at late time and affect, for instance, predictions from Big-Bang Nucleosynthesis (BBN). Whereas this possible imprint of the scalar dynamics on post-inflationary cosmology may lead to severe constraints on the inflationary model considered, it is interesting to note that having a global vision of the different phases of evolution of the Universe generally helps fixing the number of $e$-folds of inflation required to accommodate CMB measurements, leading to more precise predictions regarding the inflation observables \cite{Heurtier:2019eou,Argurio:2017joe,Heurtier:2017nwl}. In this section, we review the different stages which are inherent to such inflationary models and present the different constraints which can be set during those different epochs. Throughout this paper, we use the subscripts "end", "kin" and "eq" in order to  refer respectively to the end of inflation, the end of kination domination that represents the onset of radiation domination era, and matter-radiation equality representing the beginning of matter domination era.

\subsection{Reheating Mechanism}

The production of hot particles at the end of inflation in non-oscillatory models cannot be realized in a similar manner than in the oscillatory case. Indeed, in the latter case the mass of the inflaton field is given by its value at the minimum and the field undergoes multiple oscillations until $H\sim \Gamma_{\phi}$ when the inflaton perturbatively decays into lighter particles. In the former case, the curvature of the potential goes to zero in the limit $\phi\to\infty$, and while the field is rolling, its ability to decay is vanishing accordingly. However, successful ways to reheat the Universe exist in the literature, such as {\em instant pre-heating}~\cite{Felder:1998vq,Campos:2002yk}, {\em curvaton pre-heating}~\cite{Feng:2002nb,BuenoSanchez:2007jxm,Matsuda:2007ax} or {\em Ricci reheating}~\cite{Opferkuch:2019zbd}. Whereas those different mechanisms require a certain amount of model building, it is known that SM particles can always be produced out of the vacuum fluctuations at the end of inflation through gravitational production \cite{Ford:1986sy,Chun:2009yu}. In order to avoid adding extra ingredients to our scenario, we will consider this last possibility throughout this paper.

Following Ref.~\cite{Dimopoulos:2017zvq}, we denote with the subscript '${\rm end}$' the quantities defined at the time where the slow roll ends ($\epsilon(\phi_{\rm end})\equiv 1$). The energy of radiation produced through gravitational reheating is therefore given by
\be\label{eq:fractiongrav}
\left.\rho_{\rm grav}\right|_{t=t_{\rm end}} = \frac{g_{\star,{\rm end}} q}{1440\pi^2}\left(\frac{H_{\rm end}}{M_p}\right)^2 \rho_{\rm end}\,,
\ee
where $\rho_{\rm end}$ is the energy density at the end of inflation, $g_{\star,{\rm end}}$ is the number of relativistic degrees of freedom present in the thermal bath at the time of reheating and $q$ is an $\mathcal O(1)$ efficiency factor that we take to be equal to one throughout this paper. 
\subsection{Kination Era}
As one can see from Eq.~\eqref{eq:fractiongrav}, the fraction of energy which is injected into SM and DM particles at the end of inflation is much smaller than one. After inflation, the Universe therefore remains dominated by the inflaton energy density. However, because the period of slow-roll is over for $\varphi>\varphi_{\rm end}$, and the field simply rolls down a potential which vanishing exponentially fast, the Universe is dominated nearly entirely by the kinetic energy of the scalar field. This period of so-called {\em kination} domination is characterized by an equation of state parameter $w_{\rm kin}\approx 1$ and an energy density redshifting like $a^{-6}$. 
In the absence of any other source of energy, the equations of motion which describe the evolution of the canonically normalized scalar field are given by
\bea\label{eq:scalardyn}
\ddot{\varphi}+3H \dot{\varphi}+V_\varphi=0 \\
H=\sqrt{\dfrac{\dot{\varphi}^2}{6}+ \dfrac{V(\varphi)}{3}} 
\eea
For numerical purposes, it is convenient to define the dimensionless variables
\bea
x = \dfrac{\varphi}{M_p}\,, \hspace{0.5cm} y= \dfrac{\dot{\varphi}}{H_0 M_p} \,,
\eea
for which Eq. \ref{eq:scalardyn} splits into two first order differential equations, namely
\bea
x'= \dfrac{y}{\bar{H}}, \hspace{0.5cm}
y'= -3 y- \dfrac{\bar{V}_x}{\bar{H}},
\eea
where the prime denotes the derivative with respect to the number of e-folds $N=\ln a$, $\bar{H}\equiv \dfrac{H}{H_0}$, $\bar{V}\equiv\dfrac{V}{H_0^2M_p^2}$ and the Hubble parameter today is taken to be $H_0= 67.27\pm 0.60$ km s$^{-1}$ Mpc$^{-1}$ $=5.891\times 10^{-61} \, M_p$ \cite{Planck:2018vyg}. During inflation, it is sufficient to solve this simple system of equations. 

After inflation ends, we use Eq.~\eqref{eq:fractiongrav} to estimate the amount of energy which is produced through gravitational reheating at $\varphi=\varphi_{\rm end}$. From that point, the energy densities of matter and radiation ought to be taken into account in the Friedmann equations, giving 
\bea
\bar{H}= \sqrt{\dfrac{y^2}{6}+\dfrac{\bar{V}(x(N))+\bar{\rho}_{\rm rad}(N)+\bar{\rho}_{\rm m}(N)}{3}}\,.
\eea
In this equation, we have defined dimensionless variables for the radiation and matter energy densities:
\bea
\bar{\rho}_{\rm rad}(N)&=& 3\,  \Omega_{{\rm rad},0} \, e^{-4N}\,, \\
\bar{\rho}_{\rm m}(N)&=& 3\,  \Omega_{{\rm m},0} \, e^{-3N}\,,
\eea
with $\Omega_{{\rm rad},0}\sim 10^{-3}$, and $\Omega_{{\rm m},0}= 0.3166 \pm 0.0084$, being the values of the radiation and matter density parameters \cite{Planck:2018vyg}. At $\varphi=\varphi_{\rm end}$ we impose that the energy density of particles produced gravitationally given by Eq.~\eqref{eq:fractiongrav} equals $\bar{\rho}_{\rm grav}= \bar{\rho}_{\rm m}+\bar{\rho}_{\rm rad}$ where $\bar{\rho}_{\rm grav}= {{\rho}_{\rm grav}}/({H_0^2 M_p^2})$. The total energy density is then given by $\bar{\rho}_{\rm tot}=\dfrac{y^2}{2}+\bar{V}+\bar{\rho}_{\rm rad}+\bar{\rho}_{\rm m}$ and the equation-of-state parameter of the inflaton sector $w_\varphi$ is given by 
\bea
w_\varphi(N)= \dfrac{y^2/2-\bar{V}}{y^2/2+\bar{V}}\,.
\eea
As mentioned above, the scalar evolution then proceeds as follows:
\begin{itemize}
    \item {\it Kination}: After the inflation ends the scalar field enters a phase of kination  where the inflaton kinetic energy is dominant. This corresponds to $w_\varphi=1$, $\rho_{\rm kin} \propto a^{-6}$ and $a\propto \sqrt[3]{t}$  hence $H= \dfrac{1}{3t}$. Accordingly, by solving Eq. \ref{eq:scalardyn}, we have $\varphi(N)$ during kination
\be
\varphi(N)=\varphi_{\rm kin}+\sqrt{6} \, M_p \, \left( N-N_{\rm kin}\right),
\ee
where $N_{\rm kin}$ and $\varphi_{\rm kin}$ are given at the beginning of kination.

\item {\it Radiation}:
 When SM radiation starts dominating over the scalar field energy density, the phase of kination effectively ends and the scalar field evolves in a radiation-dominated Universe. Its evolution can be described by
\bea 
\varphi_{\rm rad}=\varphi_{\rm kin}+\sqrt{\dfrac{2}{3}} \, M_p \, \ln\left( \dfrac{H_{\rm kin}}{H_{\rm rad}}\right),
\eea 
where $H_{\rm rad}^2=\dfrac{2\rho_{\rm rad}}{3M^2_p}$. 

\item {\it Matter domination}:
The matter-radiation equality happens at {$N_{\rm eq}=\log\left( \dfrac{\Omega_{{\rm rad},0}}{\Omega_{{\rm m},0}} \right)$}, where the matter energy density starts to dominate the universe.

\end{itemize}

\subsection{Late Slow-Roll and Tracking Regime}
While the Universe energy density decreases during radiation domination (during which $\rho_{\rm rad}$ drops as $a^{-4}$), there is a possibility that the scalar field enters a new regime of slow roll when $H\sim \sqrt{V''(\varphi)}$. In that case, the scalar field behaves like a new component of dark energy, whose relative abundance grows, as compared to matter and radiation. Once this energy component is closed to dominating the energy density of the Universe, this regime of slow-roll ends naturally, and the field dynamics can end up tracking the dynamics of the background energy density, whenever it is made of matter or radiation. This is typically the case if the slope of the scalar potential is still steep enough and the field cannot just behave like a dark energy component only by itself. In models of quintessential inflation, however, the slope of the potential may vanish sufficiently fast, and although the universe enters a tracking regime where $\rho_\varphi$ tracks radiation or matter for a while, it may start slow-rolling again and behave as a cosmological constant at present time $\rho_\varphi\sim \rho_\Lambda= \Lambda M_p^2 \sim 3.47 \times 10^{-121} \, M_p^4$. Tracking solutions have been discussed in the framework of quintessence models \cite{Zlatev:1998tr} as a nice explanation for two problems, namely, the fine tuning problem of the dark energy, and the cosmic coincidence problem.
Following Ref. \cite{Steinhardt:1999nw}, the criteria for a tracking solution is given by 
\bea
\Gamma := \dfrac{V'' V}{(V')^2} \geq  1 \,\,\,\,\,\, {\rm and} \,\,\,  \,\,\, 
\dfrac{d \ln(\Gamma -1)}{d\ln a}
\eea
Although we do not address the cosmological constant problem in this work, tracking regimes typically appear if the field rolls fast enough to the lower plateau of its potential, and the scalar field may quickly participate to about 20\% of the total energy budget of the Universe. Since the scalar field does not interact with SM particles, such a contribution during radiation domination corresponds to dark radiation, and can have a dramatic impact on the the dynamics of BBN or the emission of the CMB spectrum. 
\begin{figure*}
    \centering
    \includegraphics[width=0.48\linewidth]{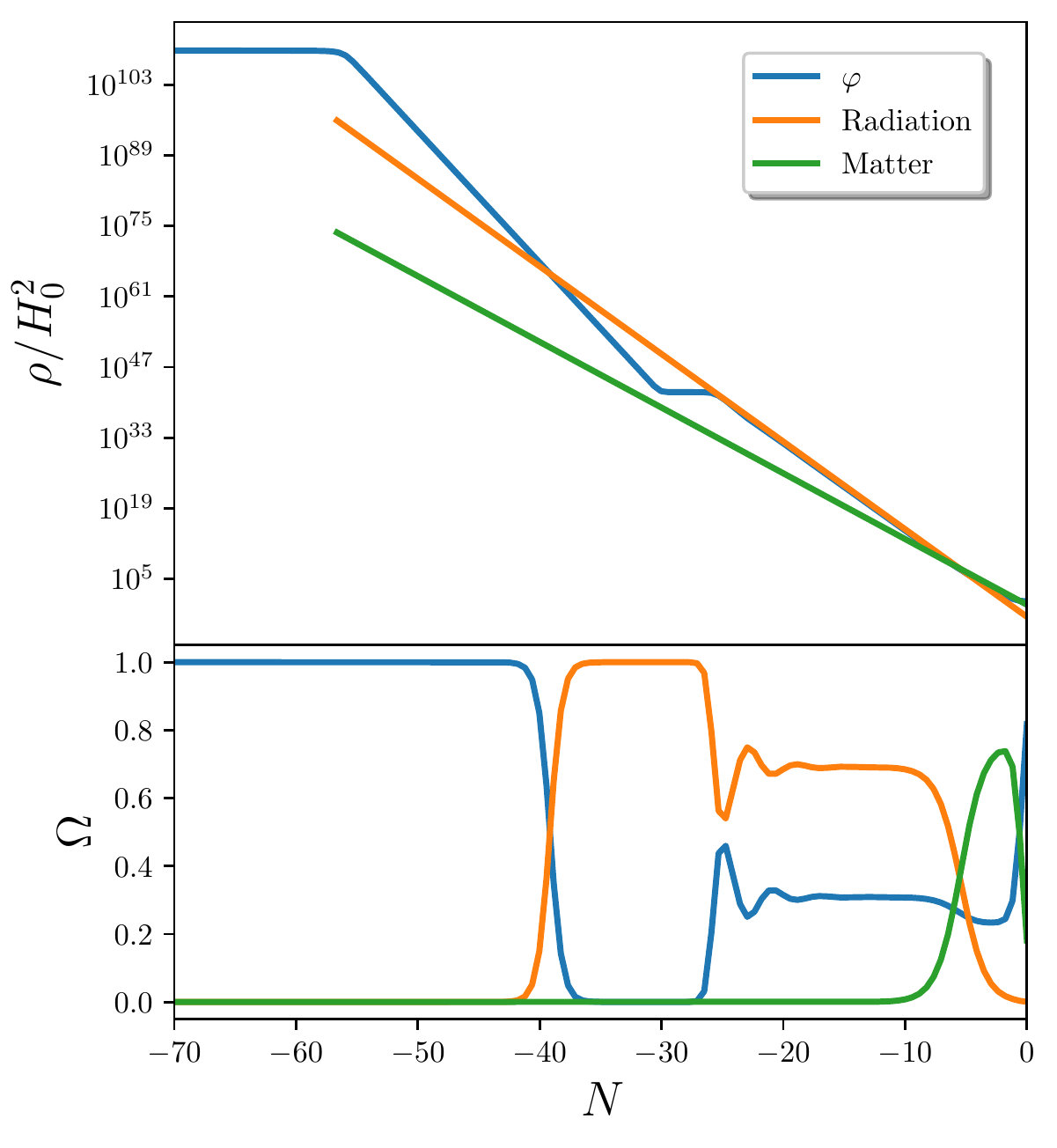}\hspace{0.04\linewidth}\includegraphics[width=0.48\linewidth]{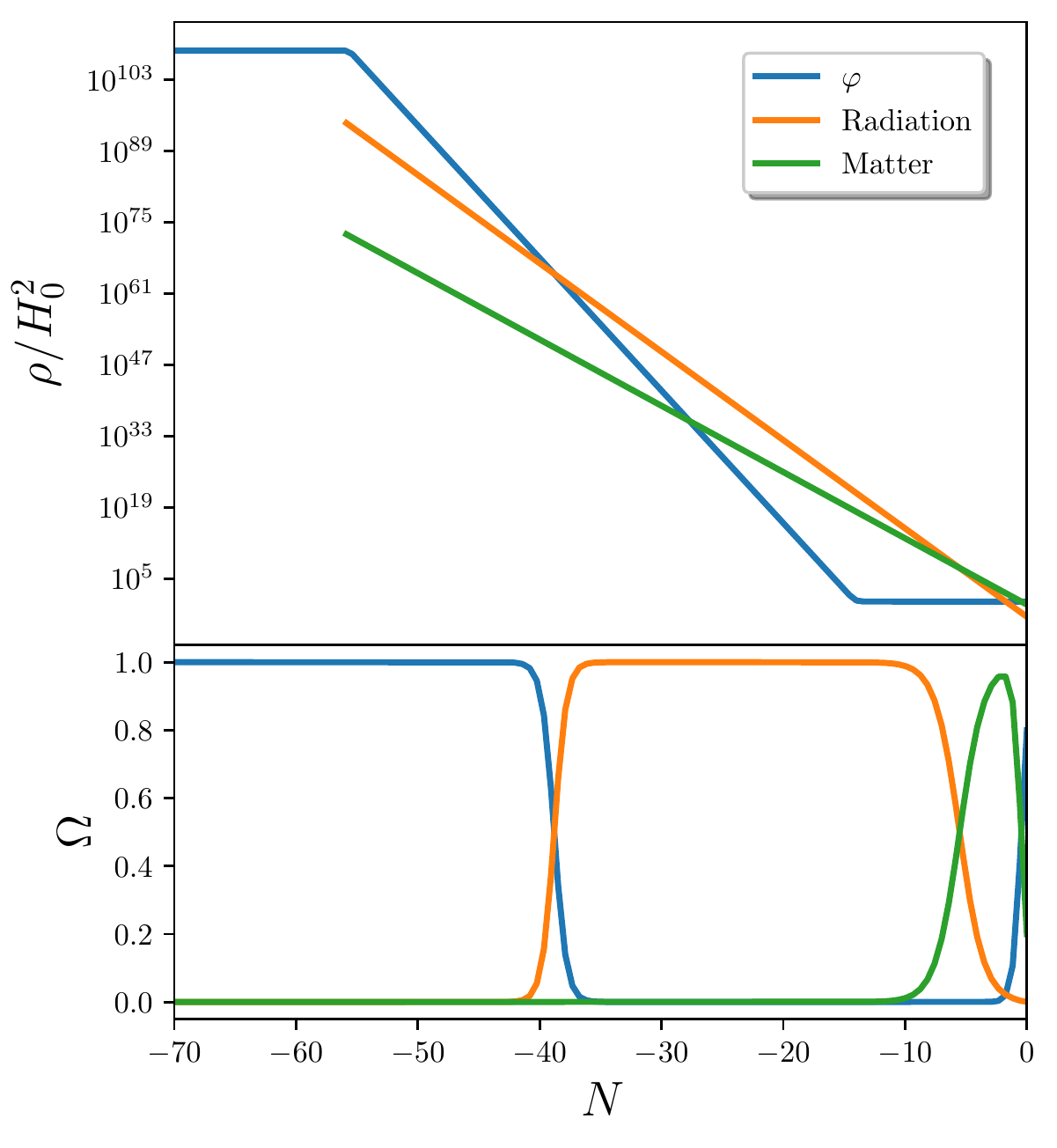}
    \caption{ \label{fig:tracking} \footnotesize Illustration of models of non-oscillatory inflation that lead to a tracking regime (left panel) or behave like kination until the domination of the cosmological constant arises. The Model II is used in these figures with $f_0=5\times 10^{-8}$, $\Lambda = 0.1 M_p$, $b=0.9$ (left panel), and $b=3.5$ (right panel). }
\end{figure*}
In Fig.~\eqref{fig:tracking}, we illustrate two situations in which the scalar field starts slow rolling early (left panel), and reaches a tracking regime during the radiation dominated era, or slow rolls only later (right panel) either at 0 or at small positive values if the potential is added a cosmological constant.
\subsection{Scalar Field Dynamics}
When solving numerically the equations of motion described above, one needs to ensure in the one hand that the scalar sector is well described by a single-field dynamics, and in the other hand that the field does not end up being stuck in a metastable minimum.

Along the scalar field trajectory, it is important to keep in mind that the sgoldstino is assumed to track its potential minimum. However, there may exist situations in which the mass of the sgoldstino may become comparable or smaller than the Hubble parameter. In that case, it is expected that the oscillations of the sgoldstino may become sizeable, which would alter the post-inflationary dynamics. In Sec.~\eqref{sec:SUSYB-QI} we have derived the value of the sgoldstino mass as a function of the inflaton scalar field. In our simulations, we will restrict ourselves to the cases where this mass remains larger than the Hubble parameter at all time during the cosmological history. 

Furthermore, we have seen that a large SUSY breaking scale can produce a bump along the inflation potential with various locations and bumpiness, depending on the different parameters of the model, like the UV cutoff scale or the $\alpha/b$ parameter. In our simulation, we eliminate all the points which would lead the inflaton to be blocked by such a bump.

\subsection{Inflation Observables}
In order to compute the inflation observables, as measured from the CMB spectrum, we need to control how many $e$-folds of inflation are required since the CMB pivot mode measured by Planck exited the horizon until the end of inflation, given the post-inflationary history of the Universe that we consider.
Cosmological scales exit the horizon at $N_*$ that is given by the relation ~\cite{Dimopoulos:2017zvq} 
\bea
k=a_\star H_\star \ \Leftrightarrow\ e^{N_\star} = 2 \dfrac{H_\star}{H_k} \left( \dfrac{a_{\rm end}}{a_{\rm kin}} \right)  \left( \dfrac{a_{\rm kin}}{a_{\rm eq}} \right)  \left( \dfrac{a_{\rm eq}}{a_{k}} \right),
\eea
where $k= 0.05 \, {\rm Mpc}^{-1}$ is the pivot scale and $a_k$ is the value of the scale factor at the time of horizon reentry. This equation can then be re-written under the form
\be
N_\star = -\ln \frac{k}{a_0H_0}+\frac{1}{6}\ln\frac{\rho_{\rm kin}}{\rho_{\rm end}}+\frac{1}{3}\ln\frac{\rho_{\rm eq}}{\rho_{\rm kin}}+\ln\frac{H_\star}{H_{\rm eq}}+\ln\frac{a_{\rm eq}H_{\rm eq}}{a_0 H_0}\,,
\ee
which is what we used in our simulations to determine the value of $N_\star$ required for each point in the parameter space.

\begin{figure*}
    \centering
    \includegraphics[width=\linewidth]{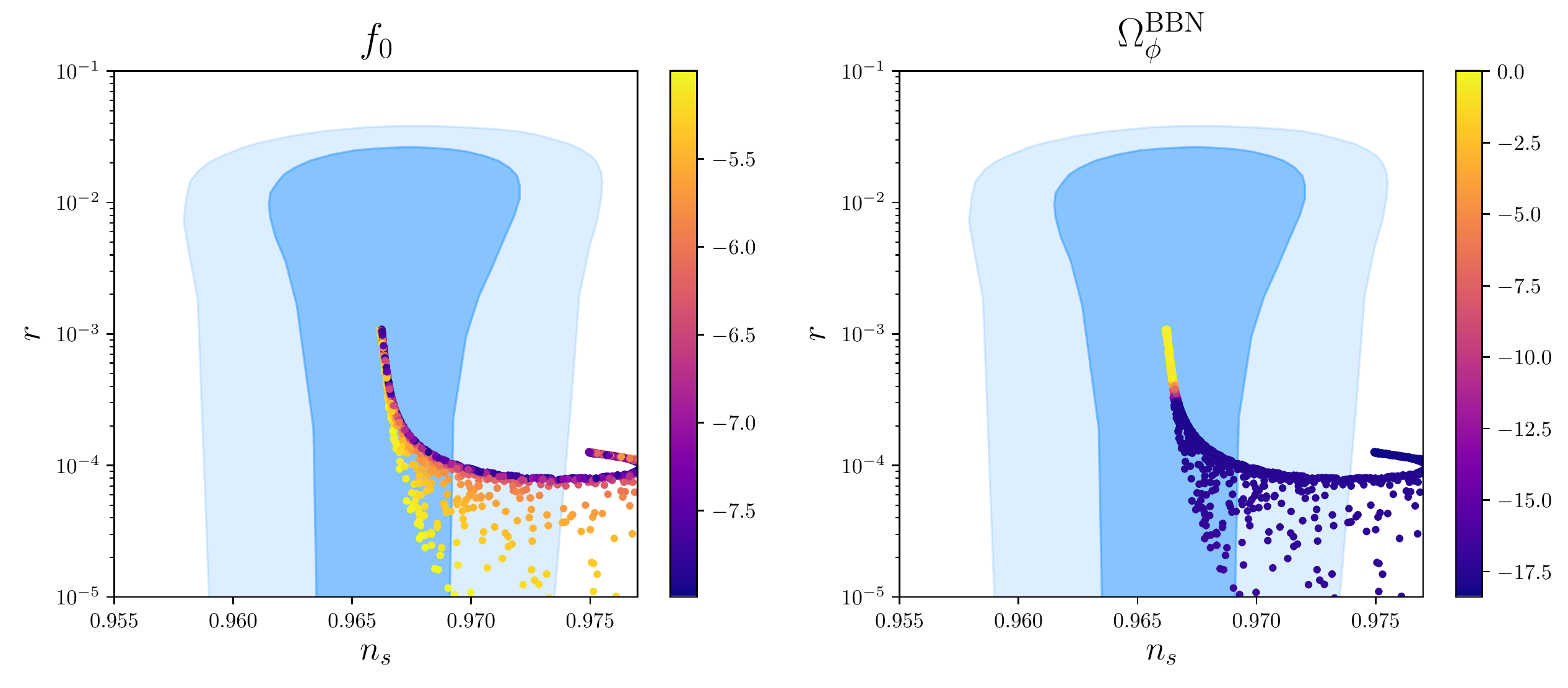}
    \caption{\label{fig:obs_I} \footnotesize Predictions for the inflation observables in the case of Model I. The left panel indicates in the color bar the value of $\log_{10} f_0/M_p$ whereas in the right panel the color bar indicates the value of the scalar field relative abundance during BBN.}
\end{figure*}

\begin{figure*}
    \centering
    \includegraphics[width=\linewidth]{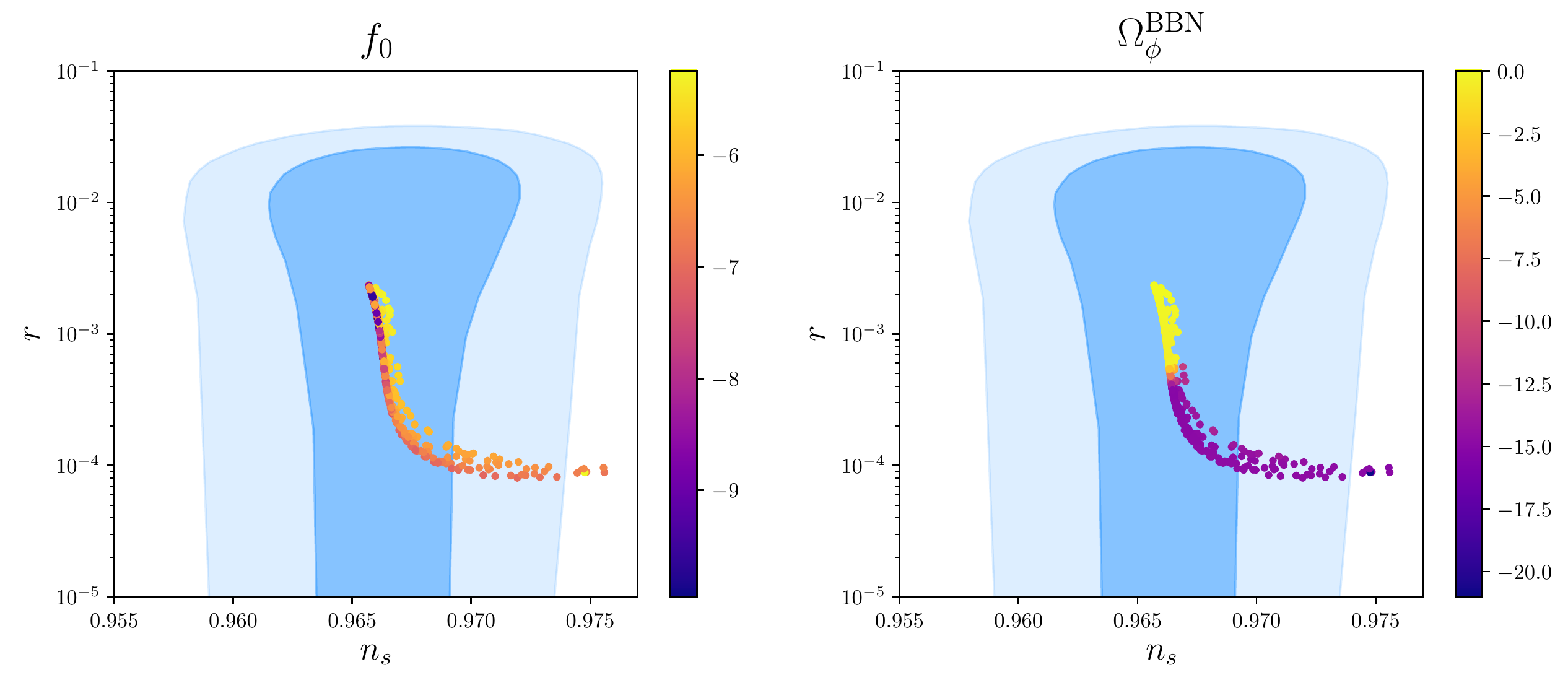}
    \caption{\label{fig:obs_II} \footnotesize Predictions for the inflation observables in the case of Model II. The left panel indicates in the color bar the value of $\log_{10} f_0/M_p$ whereas in the right panel the color bar indicates the value of the scalar field relative abundance during BBN.}
\end{figure*}

In Figs.~\ref{fig:obs_I} and \ref{fig:obs_II} we present our results for the tensor-to-scalar ratio $r$ and spectral index $n_s$ predicted in Model I and Model II, respectively. On the left, the logarithmic values of the SUSY breaking scale $f_0$ are indicated in the color bar, whereas on the right, the color bar indicates the logarithmique value of scalar relative abundance during BBN. Different points correspond to different values of the UV scale $\Lambda$ and the parameters $\alpha$ or $b$.
The results are compared to the most recent constraints combining Planck and Bicep/Kek data \cite{BICEPKeck:2022mhb}. Interestingly, demanding that no tracking regime takes place during BBN (by demanding that $\Delta N_{\rm eff}<0.284$ according to \cite{Planck:2018vyg}) rules out most of the points which are at the center of the CMB contours. Whereas in the case of model I this does not set any limit on the value of the corresponding SUSY breaking scale, in the case of Model II one can notice that this imposes a lower bound on the SUSY breaking scale $f_0\gtrsim 10^{-8}$ which suggests that a large SUSY breaking scale avoids falling in an extended tracking regime during BBN. Finally, in both models, the value of the spectral index is predicted to be larger than $n_s\gtrsim 0.966$.
In the case of Model I, another interesting feature is visible, as it appears that the smaller the SUSY breaking scale $f_0$, the larger is the tensor-to-scalar ratio $r$.

%
%

%
\section{Primordial black holes}
\label{subsec:pbh}
As one can see from Figs.~\ref{fig:potI} and \ref{fig:potII}, one of the main effect of a large SUSY breaking scale on the inflaton potential is to tilt the potential at a location that is a function of the different parameters of each model. As we have seen in the previous sections, the presence of such a bumpy region along the scalar field trajectory can lead to a significant modification of the number of $e$-folds of inflation and alter the post-inflationary dynamics of the scalar. Furthermore, it can also lead to a well-known regime of {\em ultra slow roll} {that} typically can be responsible for the formation of primordial black holes (PBHs) in the early Universe. In this section, we study this possibility and show that the presence of SUSY breaking in sgoldstino-less inflation models is able to explain the relic density of dark matter under the form of PBHs, or to form a large density of mini-PBHs {that might} dominate the energy density of the Universe before evaporating and therefore modify the post-inflationary dynamics predicted by the model.

\subsection{General Principle}

The production of a significant fraction of PBHs in the early Universe from inflation is usually achieved by introducing a phase of ultra slow-roll (USR)~\cite{Dimopoulos:2017ged} along the inflation trajectory. During ultra slow-roll, the field velocity decreases rapidly, and so does the first slow-roll parameter $\epsilon$, leading to a peak in the evolution of the second slow-roll parameter $\eta$. If the value of $\eta$ exceeds 3 during that time, the scalar-perturbation power spectrum starts to grow exponentially fast~\cite{Byrnes:2021jka}. Whereas the simplest vanilla models of inflation typically do not have any phase of USR along their trajectory, models of so-called {\em inflection-point inflation} were constructed in the literature in order to produce the correct amount of dark matter from primordial perturbations~\cite{Ballesteros:2017fsr, Karam:2022nym, Dalianis:2018frf}. In our scenario, the breaking of SUSY at a scale comparable to the inflation scale, which translates into sizeable excursions of the sgoldstino away from the origin, typically produces inflection points along the inflation potential, as we have seen in Sec.~\ref{sec:modelI} and \ref{sec:modelII}. Whereas we have seen in Sec.~\ref{sec:inflationobservables} that the presence of this inflection point can modify the value of the inflation observables, or even lead to the field to being stuck in a metastable minimum, there exist situations where the presence of this inflection point leads to an USR regime, leading to the formation of PBHs in the post-inflationary Universe. In what follows, we recall general results about the production of PBHs during a period of kination domination, and exhibit the distributions that can be produced due to the presence of SUSY breaking in our models.




\subsection{Primordial Perturbations and Power Spectrum}
Primordial black holes typically form when an overdense region of spacetime collapses when it re-enters the Hubble horizon. In order to calculate the number density of PBHs that formed in the early Universe, it is therefore necessary to track the evolution of the spacetime perturbations power spectrum and to evaluate the amount of such overdense regions at the time of horizon reentry. The evolution of the power spectrum is calculated using the Mukhanov-Sasaki equation \cite{Ballesteros:2017fsr}
\be
\frac{\text{d}^2u_k}{\text{d}N^2}+(1-\epsilon)\frac{\text{d}u_k}{\text{d}N}  + \left[\frac{k^2}{(e^NH)^2} + (1+\epsilon-\eta)(\eta-2) - \frac{\text{d}(\epsilon - \eta)}{\text{d}N} \right]u_k = 0\,,
\ee
where we use the slow-roll parameters
\be
\epsilon = \frac{1}{2}\varphi'^2, \ \ \ \ \eta = \epsilon - \frac{1}{2}\frac{\text{d}\ln \epsilon}{\text{d}N}\,,
\ee
and the $u_k$ functions denote the Fourier modes of the so-called Mukhanov-Sasaki variable $u = a\delta\varphi + a^2 \varphi'\Phi$ ($\Phi$ being the gravitational potential and $\delta\varphi$ denoting the quantum perturbation of the inflaton field). The curvature power spectrum can then be calculated after horizon exit for each $k$ mode  using the simple limit \cite{Ballesteros:2017fsr}
\be
\mathcal{P}_{\mathcal{R}}(k)  = \lim_{k \gg aH} \frac{k^3}{2\pi^2} \left|\frac{u_k}{e^N\varphi'}\right|^2 .
\ee
 Curvature perturbations turn into statistical fluctuations of the density contrast $\delta\equiv \delta \rho/\rho$. The formation of a primordial black hole typically takes place when this density contrast goes above a critical value $\delta_c(k)$ \cite{Young:2014ana}. At first-order in perturbation theory, $\delta$ follows a Gaussian centered law, and the mass fraction $\beta$ of PBHs as compared to the total mass of the Universe at the time of formation is given by
\be\label{eq:beta}
\beta(k) = 2\int_{\delta_c(k)}^{\infty} \frac{1}{\sqrt{2\pi\sigma^2(k)}} e^{-\delta^2 / 2\sigma^2(k)}\text{d}\delta\,.
\ee
In this expression, the global factor 2 is needed for normalization reasons (see e.g. Ref.~\cite{liddle2001cosmological}  and references therein) and the $k$-dependency of $\delta_c(k)$ lies implicitly in the epoch during which the $k$ mode re-enters the horizon. Indeed, different epochs have different equations of state $w = p/\rho$, and numerical simulations lead to~\cite{Harada:2013epa}
\be\label{eq:deltac}
\delta_c (k) = \frac{3(1+w)}{5+3w}\sin^2\left( \frac{\pi\sqrt{w}}{1+3w} \right) \, .
\ee
The variance $\sigma(k)$ can be calculated as
\be\label{eq:variance}
\sigma^2(k) = \int_{k_*}^{k_{end}} 4\left( \frac{1+w}{5+3w}\right)^2
\mathcal{P}_{\mathcal{R}}(q) \left(\frac{q}{k}\right)^4 e^{-q^2/k^2} T\left(\frac{q}{k\sqrt{3}}\right) \text{d}\ln q\,,
\ee
where we follow the usual procedure of applying a Gaussian window function that prevents scales vastly different from $1/k$ to contribute to the collapse of density contrasts of such a scale~\cite{Ballesteros:2017fsr}. The relation between the comoving curvature and density contrast power spectra at horizon crossing is simply a multiplication by a factor that only depends on the equation of state~\cite{Green:2004wb}. Finally, we apply a transfer function $T$ that accounts for the sub-horizon density perturbations evolution  of the form~\cite{Green:2020jor}
\be
T(x) = 3\ \frac{\sin x -x\cos x}{x^3} \,.
\ee
\subsection{Mass Fraction During Kination}
For a given post-inflationary chronology, each wavelength re-enters the horizon at a given time and may form PBHs with a given mass. In our case, when inflation ends, according to Eq.~\eqref{eq:fractiongrav} a fraction 
\be
\eta = \frac{g_{\star,{\rm end}} q}{1440\pi^2}\left(\frac{H_{\rm end}}{M_p}\right)^2\,,
\ee
of the total energy density is produced through gravitational reheating and thereafter redshifts like radiation as $\rho_{\rm rad}\propto a^{-4}$. For a kination period with equation of state parameter $w>1/3$, the period of kination domination lasts over a number of $e$-folds
\be\label{eq:efolds}
\Delta N_{\rm kin} = \left(\frac{1}{\eta}\right)^{\frac{1}{3(1+w)-4}}\,.
\ee
Denoting by $T_{\rm kin}$ the temperature of the thermal bath at the end of this period, one can write the mass (in Planck units) of a PBH that forms during the kination dominated era as a function of the comoving wavelength $1/k$ as \cite{Bhattacharya:2019bvk} 
\be\label{eq:mass}
M(k) = 4\pi\gamma\left( \frac{\pi^2 g_{*}^{\text{eq}}}{45} \right)^{\frac{1}{1+3w}} \left( \frac{g_{s}^{\text{eq}}}{g_s(T_{kin})}\right)^{\frac{3w-1}{3(3w+1)}}
T_{kin}^{-\frac{3w-1}{3w+1}} \left( \frac{a_{\text{eq}}T_{\text{eq}}}{k} \right) ^{\frac{3(1+w)}{3w+1}}\,,
\ee
where $\gamma$ is an $\mathcal O(1)$ parameter that represents the mass proportion of a Hubble patch that effectively collapses into a black hole. For simplicity we take this parameter equal to one in the upcoming simulations. For masses larger than $\gtrsim 10^{16}\mathrm{g}$, PBHs have a lifetime longer than the edge of the Universe and their fraction of the cold dark matter relic density today is \cite{Bhattacharya:2019bvk} 
\begin{multline}\label{eq:fpbh}
f_{\text{PBH}}(M) = \frac{\Omega_{\text{PBH}}(M)}{\Omega_c} \\
 = \frac{\gamma}{T_{\text{eq}}} 
 \left( \frac{g_s(T_{\rm kin})}{g_s(T_{\text{eq}})} \right)^{1/3}
 \left( \frac{\Omega_m h^2}{\Omega_c h^2} \right)
 \left( \frac{90}{\pi^2 g_*(T_{\rm kin})} \right)^{\frac{w}{1+w}}
 ( 4\pi\gamma )^{\frac{2w}{1+w}} T_{\rm kin}^{\frac{1-3w}{1+w}}
 \beta(M) M^{-\frac{2w}{1+w}}\, . 
\end{multline}
After the kination dominated era ends, the Universe begins being radiation dominated again, and the modes {that} re-enter the horizon then form with the usual mass and dark matter fraction spectrum, which can be obtained from Eqs.~\eqref{eq:mass} and \eqref{eq:fpbh} by sending $T_{\rm kin}\to T_{\rm eq}$ and $w\to 1/3$.

\subsection{Results}

Let us now present our results regarding the distributions that can typically be obtained in our models. First of all, let us note that the Model I that we considered in Sec.~\ref{sec:modelI} is different from the second in the sense that the plateau/bump generated from the breaking of SUSY is only located around positive values of $\varphi=0$. For this reason, this model tends to produce an USR period {that}, if it ever exists, only happens towards the very end of inflation. This lets very little room  to explore a significant production of PBHs, and we therefore focus our searches only on the Model II, exhibited in Sec.~\ref{sec:modelII}.
\begin{table}
\begin{center}
\begin{tabular}[c]{c c c c c} \toprule
     $f_0$ & $\beta$ & $\Lambda$ & $V_0$ & $M[\mathrm{g}]$\\ \midrule
     $6.34399\times 10^{-6}$ & $0.955$ & $0.64969999$ & $15.7\times 10^{-12}$ & $1.4\times 10^{4}$ \\ 
     $1.1190938\times 10^{-5}$& $0.84$ & $1.392$ & $3.25\times 10^{-11}$ & $5.5\times 10^{20}$\\\toprule  
\end{tabular}
\end{center}\caption{\label{tab:parameters} Parameters used to generate the two distributions presented in Figs.~\ref{fig:distri1} and \ref{fig:distri2}}
\end{table}
\begin{figure}
    \centering
    \includegraphics[width=0.5\linewidth]{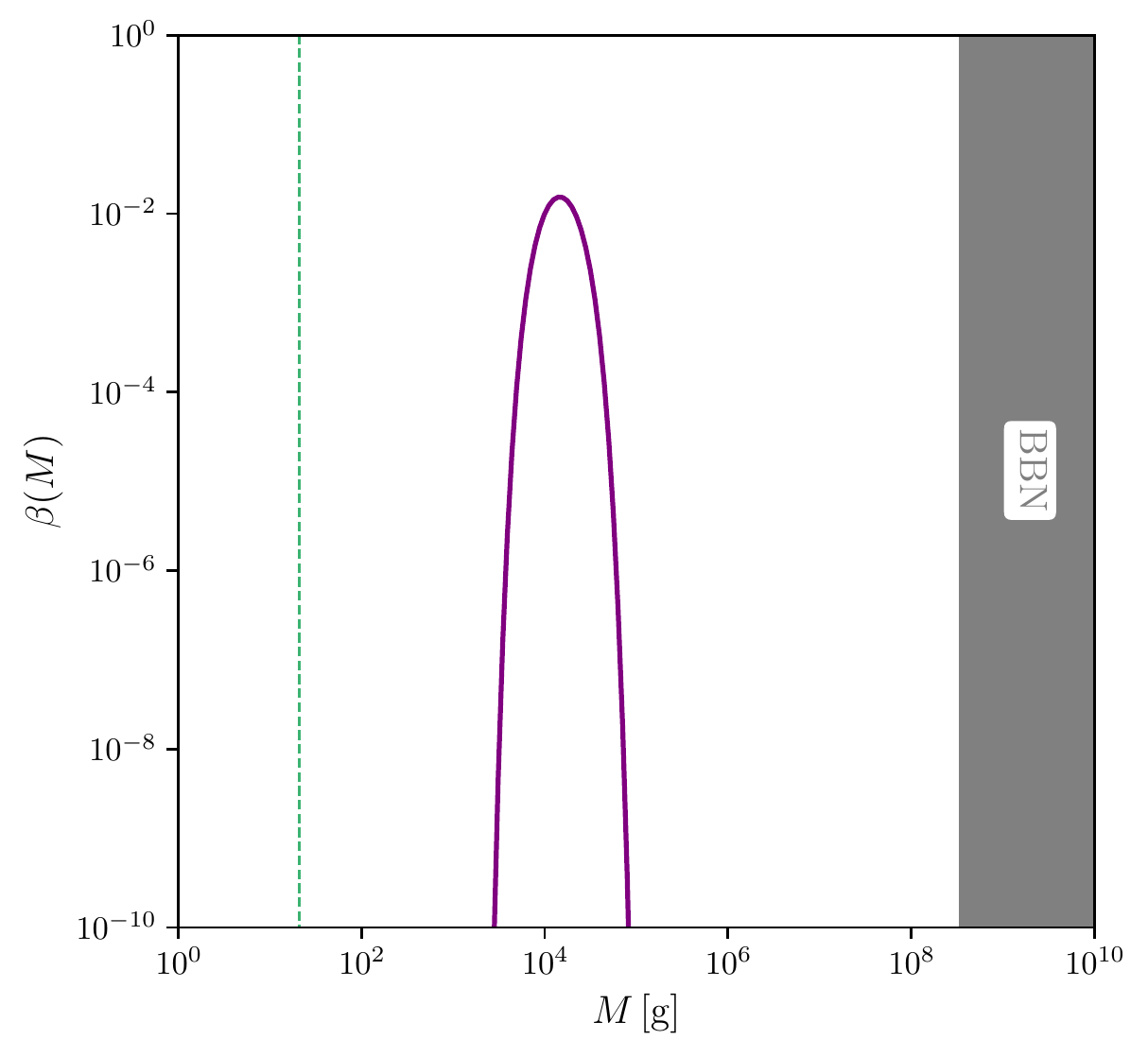}
    \caption{\label{fig:distri1}\footnotesize  Distribution of PBHs generated with the first set of parameters presented in Tab.~\ref{tab:parameters}. The gray-shaded area corresponds to PBH masses which would evaporate after during of after BBN. The green vertical line stands for masses corresponding to modes exiting the horizon at the end of inflation.}
    
\end{figure}
In the case of Model II, the location of the inflection point along the inflation trajectory and the value of the slope at the inflection point are non-trivial functions of the different parameters. Importantly, the frequency of the perturbations {that} are amplified during the USR period, as well as the height of the corresponding peak in the power spectrum {depend strongly} on {these} different factors. The value of the PBH masses {that} may be produced at horizon re-entry and their mass fraction at the time of formation are extremely sensitive on the different parameters of the model. In Tab.~\ref{tab:parameters} we provide the values of the parameters that we used for two points that we consider to be of interest. In the first case, we exhibit parameters {that} lead to an USR period at the end of inflation and therefore produce PBHs with very small masses (of order $10^4$ g in this example). In this case, the entire PBH distribution {is produced} during the kination dominated era, and one can see from Fig.~\ref{fig:distri1} that the mass fraction reaches $~\sim 0.015$ at the time of formation. With such a large energy fraction  at formation and such low masses, it is likely that these PBHs will dominate the energy density of the Universe before they evaporate. In such an extreme case, the formation of PBHs would modify the cosmological timeline that we have considered in this paper by splicing a period of early matter domination at the end of kination. This situation goes beyond the scope of this paper and will be investigated in some future work.

\begin{figure}
    \includegraphics[width=0.48\linewidth]{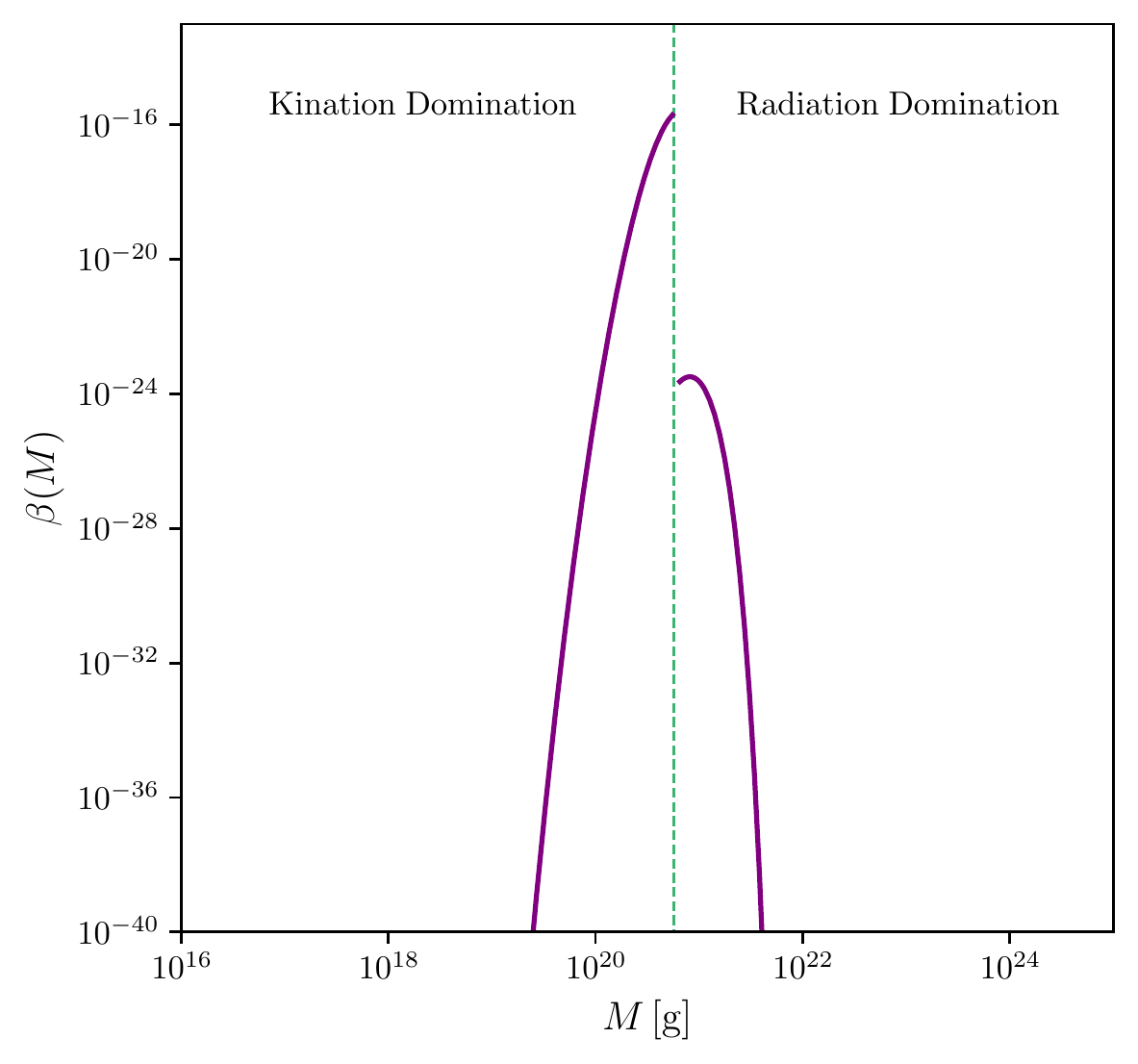}\hspace{0.04\linewidth}\includegraphics[width=0.48\linewidth]{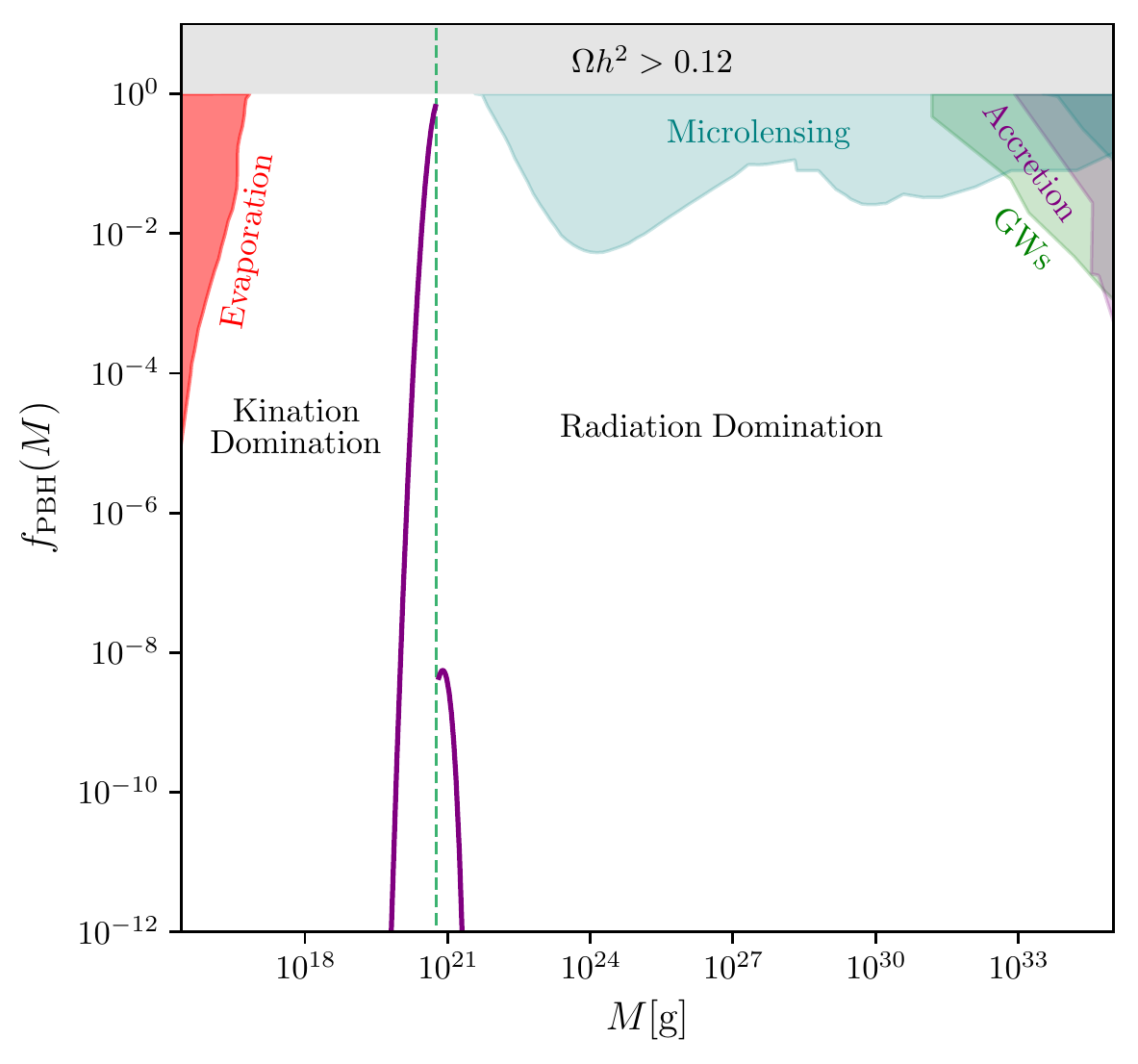}
    \caption{\label{fig:distri2}\footnotesize Mass fraction at the time of PBH formation (Left Panel) and fraction of the cold dark matter relic density under the form of PBHs (Right Panel) as a function of the PBH mass for the second set of parameters listed in Tab.~\ref{tab:parameters}. The green vertical dashed line corresponds to the mass of PBHs forming precisely at the end of the kination period. The different constraints are listed in Ref.~\cite{Green:2020jor}.}
\end{figure}

In Fig.~\ref{fig:distri2} we show the results that we obtain for the second set of parameters listed in Tab.~\ref{tab:parameters}. On the left panel we show the mass fraction of PBHs at the time of their formation, and on the right panel we present our results for their final relic density fraction. The vertical dashed line on both plots stands for the value of the PBH mass {that} is formed exactly between kination and radiation domination. Interestingly, the location of this vertical line is uniquely determined by the Hubble rate value at the end of inflation, since the fraction of energy during gravitational reheating is exclusively depending on $H_{\rm end}$ in Eq.~\eqref{eq:fractiongrav} and the duration of the kination period is uniquely given by this energy fraction in Eq.~\eqref{eq:efolds}. As we can see from the figure, this line interestingly falls right in the window which is experimentally allowed for PBHs to form the full relic density of cold dark matter in our Universe today (see e.g. Ref.~\cite{Green:2020jor} for a review of the corresponding constraints depicted in Fig.~\ref{fig:distri2}). It is interesting to note that the PBHs {that} form during the kination dominated era and those who form in the later radiation dominated era populate the Universe with very different energy fractions, which explains the sharp drop that both the mass fraction and relic density fraction feature when they cross the vertical dashed lines in Fig.~\ref{fig:distri2}. The reason for the different behavior of the distribution in the two regimes is threefold: $(i)$ the value of the critical density contrast given in Eq.~\eqref{eq:deltac} {that} is necessary for PBHs to form during kination and radiation domination are different, 
\bea
\delta_c &\approx& 0.38\qquad \mathrm{(Kination},\ w=1\mathrm{)}\,,\nonumber\\
\delta_c &\approx& 0.41\qquad \mathrm{(Radiation},\ w=1/3\mathrm{)}\,,
\eea
which means that PBHs are easier to form during kination domination than radiation domination, and $(ii)$ the evolution of the mass density into the relic density fraction depends on the Universe's evolution after the PBH formation, as can be seen from Eq.~\eqref{eq:fpbh}. Finally, {$(iii)$ the relation between the contrast and curvature spectra, which can be seen from the $w$-dependent coefficient in Eq.~\eqref{eq:variance}  (0.062 vs 0.050) also favours kination}. In the example we have chosen, the fact that PBHs have formed both during the kination and radiation dominated era can be seen from the peculiar form of the distribution, which may be one day ruled out or confirmed experimentally, if such a distribution is ever measured in the future.

Before we conclude, we should finally discuss the value of the inflation observables {that} correspond to those different examples. For the first point exhibited in Tab.~\ref{tab:parameters}, we find inflation observables with values
\be
r\approx 0.00096\qquad\text{and}\qquad n_s\approx 0.969\,.
\ee
These observables are perfectly consistent with the most recent limits set on the CMB tensor modes~\cite{BICEPKeck:2022mhb}. However, we point out the fact that the observables of the second point, which could in principle produce good PBH dark matter candidates, features a spectral index {that} is below the current experimental limits. This feature of the PBH dark matter production from inflationary models was already mentioned in previous work \cite{Ballesteros:2017fsr}. Accommodating the CMB observables while producing the correct relic density of PBHs at present time would require an additional tuning, which is out of the scope of this paper.

\section{Conclusion}
In this paper, we have studied the effect of SUSY breaking on the inflationary trajectory of sgoldstino-less non-oscillatory inflation models. We have shown that a large value of the SUSY breaking scale $f_0$ can typically bend the inflation potential around the end of inflation which can alter significantly the post-inflationary dynamics and therefore leave visible imprints in the CMB spectrum. 
Considering a purely gravitational reheating, we have explored how certain regions of the parameter space can lead to periods of {\em tracking} where the scalar field energy density tracks the energy density of the background and can contribute to about 20\% of the energy density during the radiation dominated era. Such a large contribution to the total energy content of the Universe corresponds to a dark radiation component {that} is strongly excluded if the tracking takes place during BBN. We also discussed the stability of the sgoldstino during the whole cosmological history and scanned over the parameter space in order to visualise the effect of a large SUSY breaking scale on the CMB observables given the current limits on $\Delta N_{\rm eff}$.
Finally, we have explored the possibility for the effect of SUSY breaking on the inflationary dynamics to lead to a significant production of primordial black holes in the early Universe. We have exhibited two examples where we showed that our models are able to produce PBHs {that} may either be light and come to dominate the energy density of the Universe before they evaporate in the cosmos at early time, or be heavy enough to survive at present time and constitute a large fraction of dark matter today. To put it in a nutshell, it is clear that the presence of SUSY breaking at very large energy can affect not only the dynamics of inflation, but also the subsequent scalar dynamics {that} may lead to different cosmological imprints, even if the sgoldstino remains stabilized during the whole cosmological history. Furthermore, similarly to previous attempts to unify the description of cosmology in particle physics, it is manifest in this work that a full knowledge of the Universe's evolution, from the time of inflation and cosmic reheating to present time leads to a one-to-one connection between the parameters of the model considered and observational predictions regarding the CMB. 

Although the discussion of non-oscillatory inflation in the context of sgoldstino-less models and the corresponding cosmological constraints remains largely model dependent, we believe that this work identified new directions in order to constrain experimentally the post-inflationary history of our Universe. We insist on the fact that there exist many other directions {that} could be explored in this context. It was demonstrated for example in Ref.~\cite{Goshal22XX, Opferkuch:2019zbd} that the spectrum of gravitational waves produced during inflation may lead to observable signals for gravitational wave detectors when the Universe undergoes a long phase of kination after inflation. Moreover, it is known that the production of primordial black holes {usually goes with} a secondary spectrum of gravitational waves coming from the next-to-leading order terms in perturbation theory ~\cite{Balaji:2022dbi}. Even if primordial black holes are produced in subdominant quantities, {it may be possible for} such a spectrum of induced gravitational waves to be visible by future detectors. The production of gravitational waves in the context of sgoldstino-less models of non-inflationary inflation therefore stands out as an exciting perspective that we will address in a dedicated study in the future.

\acknowledgments
The authors would like to thank particularly M. Pierre for helpful suggestions and early participation to this project, but also C. Byrnes, E. Dudas, and K. Olive for interesting discussions during the realization of this work.
The work of LH is funded by the UK Science and Technology Facilities Council (STFC) under grant ST/P001246/1. The work of AM is  supported in part by the Science, Technology and Innovation Funding Authority (STDF) under grant No. 33495.

\bibliographystyle{apsrev4-1}
\bibliography{main.bib}

\end{document}